\documentclass{article}
\usepackage{graphicx} % Required for inserting images
\usepackage{amsmath}
\usepackage{amsfonts}
\usepackage{xcolor}
\usepackage{algpseudocode}
\usepackage{algorithm}
\usepackage{bm}
\usepackage{ulem}  % strikethrough text
\usepackage[colorlinks=true, allcolors=black]{hyperref}
\usepackage{cleveref}\usepackage[T1]{fontenc}
\usepackage[
url = false,
isbn=false,
backend = biber,
style = numeric,
sorting = nty
]{biblatex}
\usepackage{setspace}
\usepackage{lineno}

\renewbibmacro{in:}{}
\title{Unbiased likelihood estimation of the Langevin diffusion for animal movement modelling}
\author{Martin Emil Pettersen \and Ron R. Togunov \and S. Knutsen Furset \and Robert B. O'Hara}
\date{}

\begin{document}

\begin{titlepage}
    \centering
    \vspace*{5cm}
    
    {\Large \textbf{Unbiased likelihood estimation of the Langevin diffusion for animal movement modelling}} \\[1.5cm]
    {\large Martin E. Pettersen$^1$, Ron R. Togunov$^1$, S. Knutsen Furset$^1$, Robert B. O'Hara$^1$}

\textit{$^1$Department of Mathematical Sciences, NTNU, Trondheim, N-7491, Norway}\\[1.0cm]

    \raggedright{\textbf{Data availability statement.} 
    Supporting code is available at \url{https://zenodo.org/records/20037956}. The Steller sea lion data used in the case study is available at \url{https://doi.org/10.5281/zenodo.1134893} (Wilson et al., 2018).}\\[1.0cm]

    \raggedright{\textbf{Acknowledgements}
    R.R.T is supported by the Research Council of Norway (GreenPlan, Project No. 326979).
    }\\[1.0cm]
    
    \raggedright {\textbf{Author contributions.}
    MEP, RRT, and SKF designed the study and conducted the analysis. MEP and RRT prepared figures and tables. MEP, RRT, and SKF wrote the manuscript with contributions from RBO. All authors contributed ideas and helped revise the manuscript.}\\[1.0cm]

\raggedright \textit{Corresponding Author:} Ron R. Togunov \endgraf
\textit{Address:} Department of Mathematical Sciences, NTNU, Trondheim, N-7491, Norway\endgraf
\textit{Email:} ron.togunov@ntnu.no \endgraf

\end{titlepage}

%\doublespacing
%\linenumbers

\maketitle

\begin{abstract}
\begin{enumerate}
    \item An ongoing challenge in animal ecology is developing movement models that account for the autocorrelation, and often temporal irregularity, in telemetry data. Continuous-time Langevin diffusion models have been proposed to model temporally autocorrelated and irregularly sampled data. However, current estimation techniques obtain increasingly biased parameter estimates as the time between observations increases. 
    \item In this paper, we propose using Brownian bridges in an importance sampling scheme to improve the likelihood approximation of the Langevin diffusion model. In a series of simulation studies, we showed that our approach effectively removed the bias under various scenarios. We found that the precision of the estimated habitat coefficients increased for data spanning a longer duration at a lower frequency than for shorter, more frequently sampled tracks. This suggests that the model may be well suited for modelling tracking data sampled at a coarser resolution, as is common in datasets collected with older generations of animal tags. 
    \item We illustrated the application of our model using tracking data from Steller sea lions, \textit{Eumetopias jubatus}. We found that the coefficient estimates converged to values significantly different than those estimated in previous studies, suggesting that bias in conventional estimation methods may meaningfully affect ecological conclusions about habitat preference.
    \item Together, these improvements broaden the applicability of Langevin diffusion models, thereby improving ecological insight into habitat selection.
\end{enumerate}
\end{abstract}

%\subsection{Todo}

%\begin{enumerate}
   
%\end{enumerate}


\section{Introduction}
A central challenge in ecology is the estimation of the spatial distribution of a individuals of a species, known as the utilisation distribution (UD) \cite{wortonKernel1989, manly2002resource, austinSpatial2002, millspaughAnalysis2006}. Accurately quantifying this distribution is critical for informing management and conservation. Assuming abundance of individuals is proportional to habitat suitability, estimations of species-habitat association in environmental space can be used to predict the UD in geographic space \cite{boyceRelating1999, manly2002resource}. The resource selection function (RSF) is a weighting function that predicts the intensity of use given presence of certain values of environmental covariates, where \textit{species intensity} is the expected number of observations per unit area \cite{boyceRelating1999, manly2002resource, leleWeighted2006, Florko2025}. Most approaches estimate the RSF by modelling observed species location data with corresponding habitat characteristics. 

When the locations of observed individuals can be assumed to be independent, the RSF can be estimated using use-availability designs with logistic regression \cite{johnsonResource2006, leleWeighted2006} or inhomogeneous Poisson point process models \cite{millspaughAnalysis2006, aartsComparative2012, fithianFinitesample2012}. However, with the proliferation of animal tracking data, which generate serially auto-correlated data, there has been a need to develop models that account for both habitat selection (i.e., RSF) and the movement processes \cite{rhodesSpatially2005, Florko2025}. 

Location data obtained at fixed intervals can be modelled using the step selection function (SSF), which models movement between consecutive locations as a product of habitat selection and movement kernels \cite{rhodesSpatially2005, foresterAccounting2009, avgarIntegrated2016, Florko2025}. However, SSFs are scale dependent, so that the parameter estimates depend on the sampling intervals and reflect different ecological processes \cite{signerEstimating2017, fiebergHow2021}. Further, these models are less suitable for data where the sampling interval is irregular, for example, locations obtained using Argos-linked tags, tags that are duty cycled (i.e., programmed to cycle through different sampling rates), or when jointly modelling tags with different sampling intervals \cite{avgarIntegrated2016, fiebergHow2021, Florko2025}. To contend with these problems, various continuous-time models have been proposed, including SSFs with time-dependent distributions \cite{eisaguirreRayleigh2024, hofmannMethods2024}, continuous-time correlated random walks \cite{johnson2008}, move-persistence models \cite{jonsenMovement2019}, and more recently, Langevin diffusions \cite{Michelot2019}. 

The Langevin diffusion model is particularly well suited to studying habitat selection, as it explicitly models movement as a stochastic diffusion process whose long-term behaviour tends towards the UD \cite{Michelot2019}. However, current methods for statistical inference generate biased parameter estimates, particularly at lower sampling rates \cite{Michelot2019}. This problem arises because current methods for inference are based on an Euler-Maruyama discretisation of the Langevin diffusion, which increasingly flattens the target UD as the step size increases, thereby underestimating habitat selection coefficients. This causes the inference to see a much flatter UD than the truth, causing a bias towards 0 in the parameter estimates. We propose to circumvent this by generating higher-resolution integration points between measurements, allowing us to a apply an Euler-Maruyama discretisation with finer step size. Specifically, following the approach described by Durham \& Gallant \cite{Durham2002}, our inference method generates numerous higher-resolution Brownian bridges as proposal tracks between observed locations and then approximate the likelihood using these simulated tracks as proposals in a Monte-Carlo integration with importance sampling. We call this method the Brownian bridge with importance sampling (BBIS) likelihood approximation. We first introduce the Langevin diffusion model and the BBIS method. We then explore the precision and accuracy of the BBIS parameter estimates through four simulation studies that investigate the effect of sampling duration, sampling rate, temporal resolution of the bridges, and number of Brownian bridges. Finally, we also illustrated the BBIS method in a replication of the case study of the Steller sea lion, \textit{Eumetopias jubatus}, data in Michelot et al. \cite{Michelot2019} to investigate the effect of the BBIS method on coefficient estimates to in an ecological dataset.

\section{The Langevin diffusion model}

The overdamped Langevin Itô diffusion on $\mathbb{R}^2$ is given by
\begin{align} \label{eq:langevin-diffusion}
\begin{cases}
    \text{d} X(t) - \frac{\gamma^2}{2} \nabla \log(\pi(X(t))) \text{d}t = \gamma \text{d}W(t) \\
    X(0) = x_0
\end{cases} \, ,
\end{align}
where $\gamma \in (0, \infty)$ is a scaling parameter, $x_0 \in \mathbb{R}^2$ is the initial position, and $\pi: \mathbb{R}^2 \rightarrow \mathbb{R}$ is the target probability density for the diffusion process. The Langevin diffusion (\ref{eq:langevin-diffusion}) is a continuous-time Markov process and under suitable conditions on $\pi$, it has a limiting distribution with probability density $\pi$, see Roberts \& Tweedie \cite{Roberts1996} for details. Intuitively, this implies that $X$ spends more time in regions were $\pi$ is large compared to regions were $\pi$ is small. When we use (\ref{eq:langevin-diffusion}) as a model for animal movement, it is natural to identify the limiting distribution of (\ref{eq:langevin-diffusion}) with the UD by equating $\pi$ with a resource selection function (RSF). Typically the RSF is linked to spatial covariates. If we imagine $J$ spatial covariates $c_1(x), c_2(x), ..., c_J(x)$, with corresponding parameters $\beta_1, \beta_2, ..., \beta_J$, we can parametrise the RSF as 
\begin{equation*}
\pi(x) \propto \exp \left( \sum_{m = 1}^J \beta_m c_m(x) \right) \, .
\end{equation*}
We assume that we have observations of the animals location at discrete times $t_0 < t_1 < ... < t_n$, possibly with different step-sizes. We denote these observations by $x_i := X(t_i)$. In this work we assume that these observations are made without measurement error. Based on these observations we aim to use maximum likelihood estimation to infer the parameters $\gamma$, and $ \Vec{\beta} := (\beta_1, \beta_2, ..., \beta_J)$. We denote the likelihood function of the observations by $\Pi$, and the log-likelihood by $\ell$. Further denote by $p_{y, \Delta t}(x)$ the density of $X(\Delta t)$ conditional on $X(0) = y$. We can then exploit the Markov property of Itô diffusions to decompose $\Pi$ as 
\begin{equation} \label{eq:likelihood-decomposition}
    \Pi(\gamma, \Vec{\beta}) = \pi(x_0) \cdot \prod_{i = 1}^n p_{x_{i - 1}, t_i - t_{i - 1}}(x_{i}) \, ,
\end{equation} 
or equivalently 
\begin{equation} 
    \ell(\gamma, \Vec{\beta}) = \log(\Pi(\gamma, \Vec{\beta})) = \log(\pi(x_0)) + \sum_{i = 1}^n \log(p_{x_{i - 1}, t_i - t_{i - 1}}(x_{i})) \, .
\end{equation}
Michelot, et al \cite{Michelot2019}, performed maximum likelihood estimation by approximating the likelihood using an Euler-Maruyama discretisation of (\ref{eq:langevin-diffusion}). This essentially amounts to approximating the distribution of the increments $x_i - x_{i - 1}$ of the observations by a Gaussian distribution, or more succinctly
\begin{equation} \label{eq:Euler-Maruyama approximation}
p_{x_{i - 1}, t_i - t_{i - 1}}(x_{i}) \approx \frac{1}{2 \pi \gamma^2}\text{e}^{- \frac{1}{2 \gamma^2} \left|x_{i} - x_{i - 1} - \frac{\gamma(t_i - t_{i - 1})}{2} \nabla \log(\pi(x_{i - 1})) \right|^2} \, .
\end{equation}
This approximation produces large biases in the maximum likelihood estimates as the gap between observations grows larger \cite{Michelot2019}. However, we claim that this bias is not inherent to the Langevin model, but arises from the Euler-Maruyama discretisation. 

\section{The BBIS approximation} \label{sec:bbis}

\begin{figure}[H]
    \centering    \includegraphics[width=1\linewidth]{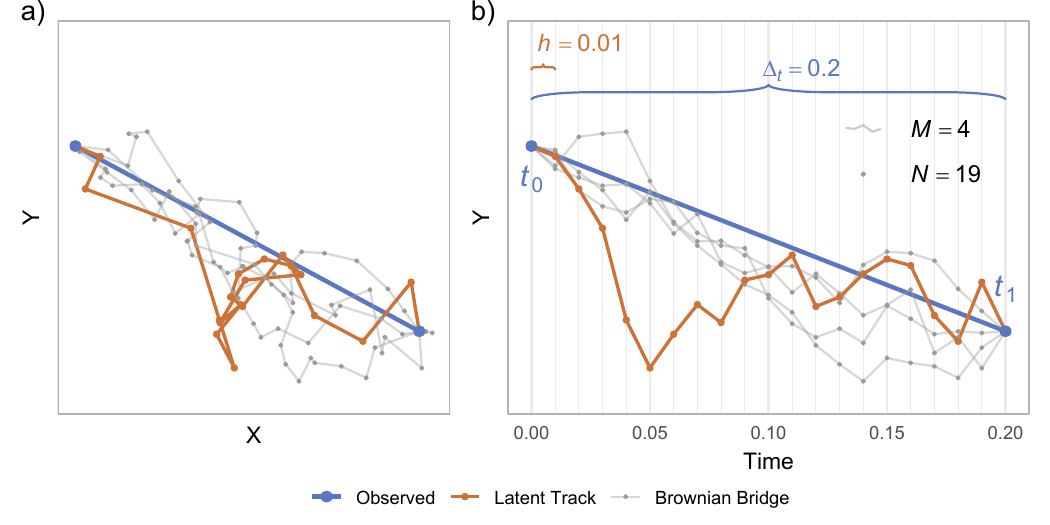}
    \caption{Illustration of the proposed BBIS for one step. Panel a) shows tracks in geographic space, and panel b) shows the y-axis coordinates of the same tracks with respect to time. Latent track based on langevin diffusion is shown in orange for 20 steps at a time interval of $h=0.01$, while observed locations (blue) are obtained at a sampling interval $\Delta_t = 0.2$. Importance sampling is here illustrated with $M = 4$ Brownian bridges composed of $N = 19$ nodes shown in grey.}
    \label{fig:concept}
\end{figure}

Instead of directly applying the Euler-Maruyama scheme, we propose an importance sampling approximation of $p_{x_0, t}(x)$ using Brownian bridges, based on the method described by Durham and Gallant \cite{Durham2002}. Essentially, this approach simulates multiple intermediate paths between consecutive observations using Brownian bridges, and approximates the transition density by averaging the Euler-Maruyama likelihood over these proposed paths. We imagine a subgrid of $N$ importance sampling nodes between $t_{i - 1}$ and $t_i$, with step size $h := \frac{t_i - t_{i - 1}}{N + 1}$. We draw $M$ random $N$-vectors $\Vec{y}_1, ..., \Vec{y}_M$ from a proposal distribution $q(\Vec{y})$. Our proposed Monte-Carlo approximation of $p_{x_{i - 1}, t_i - t_{i - 1}}(x_{i})$ is then
\begin{equation} \label{eq:monte-carlo-approximation}
    p_{x_{i - 1}, t_i - t_{i - 1}}(x_{i}) \approx \frac{1}{M} \sum_{k = 1}^M \frac{p_{x_{i - 1}, h}(y_k^{(1)}) p_{y_k^{(N)}, h}(x_i) \prod_{j = 2}^N p_{y_k^{(j - 1)}, h}(y_k^{(j)})}{q(\Vec{y}_k)} \, .
\end{equation}
We then approximate $p_{x_{i - 1}, h}(y_k^{(1)})$, $p_{y_k^{(N)}, h}(x_i)$, and $p_{y_k^{(j - 1)}, h}(y_k^{(j)})$ using the Euler-Maruyama approximation (\ref{eq:Euler-Maruyama approximation}). The Euler-Maruyama approximation still introduces some bias, but now this bias is decoupled from the step size in the observations and can therefore be made arbitrarily small by reducing the step size $h$ between importance sampling nodes. This Monte-Carlo estimator is then asymptotically unbiased as $h \rightarrow 0$, and converges in probability with rate $M^{-1}$, see Durham \& Gallant \cite{Durham2002} for details. An illustration of the BBIS approximation of the Langevin process for one time step is provided in Figure \ref{fig:concept}.

\subsection{Choice of importance sampling distribution $q$}
We use as importance sampler a Brownian bridge connecting $x_{i - 1}$ to $x_i$. Specifically we  choose as proposal distribution 
\begin{equation} \label{eq:brownian-bridge-distribution}
    q(\Vec{y}) = \phi(\Vec{y}, \mu, \Sigma) \,
\end{equation}
where $\phi(\cdot, \mu, \Sigma)$ is the multivariate Gaussian probability density function, $\mu$ is an $N$-vector defined by
\begin{equation*}
\mu_j := x_{i - 1} + \frac{j}{N + 1}(x_{i} - x_{i - 1}) \, , \quad j = 1,...,N \, ,
\end{equation*}
and $\Sigma$ is an $N \times N$-matrix defined by
\begin{equation*}\Sigma_{j,k} := \sigma^2 \cdot \left( \min(jh, kh) - \frac{jkh}{N + 1} \right) \, , \quad j,k = 1,...,N \, .\end{equation*}
The Brownian bridges can be simulated directly using standard methods for the simulation of multivariate Gaussian distributions, or using iterative methods. In this work we choose $\sigma = \gamma$, motivated by the intuition that this should make the bridges more similar to (\ref{eq:langevin-diffusion}), reducing the importance sampling error. However, it is possible that there exists more optimal choices of $\sigma$.
%Since both $q$ and the (Euler-approximated) target distribution is Gaussian, we know that the tails of $q$ are at least as heavy as those of the target distribution. This means that $q$ is a valid choice of importance sampler.

\subsection{Implementation of the likelihood}

An algorithmic description of the computation of the likelihood approximation using Brownian bridge importance sampling (BBIS) can be found in Algorithm \ref{al:likelihood-computation}. The notation $\mathcal{I}_2$ refers to the $2 \times 2$ identity matrix and $\phi(x, \mu, \Sigma)$ refers to the density of a multivariate Gaussian distribution with mean $\mu$ and covariance matrix $\Sigma$ evaluated at $x$. In order to reduce the computational load and improve the stability of gradient based optimisation we pre-simulate $M$ standard Brownian bridges from $(0,0)$ to $(0,0)$ with variance $\sigma^2 = 1$, which we then use in all subsequent likelihood evaluations. When we evaluate the likelihood for a step in the data we add to each bridge the interpolation line between the measurements and scale it to variance $\gamma^2$. This can be seen in Line \ref{li:bb-scaling}. Note also that the step in Line \ref{li:gradient-computation} requires the computation or numerical approximation of the gradients of the covariates $c_m$. This step can use a lot of compute, depending on how the covariates are specified. In most applications the gradients will have to be numerically approximated.
    
\begin{algorithm}[H]
  \caption{The BBIS likelihood approximation}\label{al:likelihood-computation}
  \begin{algorithmic}[1]
    \State Simulate $B_{j,k} \sim q$ 
    \Procedure{loglik}{$\beta_1, \cdots, \beta_J, \gamma$}
      \State $\ell \gets 0$
      \State $x_{i, j, k} \gets \frac{j}{N + 1} x_{i} + \frac{N + 1 - j}{N + 1}x_{i + 1} + \gamma B_{j,k}$ \label{li:bb-scaling}
      \State $g_{i,j,k} \gets \sum_{m = 1}^J \beta_m \nabla c_m(x_{i,j,k})$ \label{li:gradient-computation}
      \For{$0 \leq i \leq n - 1$}
        \State $L_i \gets 0$
        \For{$1 \leq j \leq M$}
            \State $L_{i,j} \gets \phi(x_{i,j,1} - x_i - \frac{\gamma^2 h}{N + 1} g_{i,j,0}, 0, \frac{\gamma^2 h}{N + 1} \mathcal{I}_2)$
            \State $L_{i,j} \gets L_{i,j} * \phi(x_{i + 1} - x_{i,j,N} - \frac{\gamma^2 h}{N + 1} g_{i,j,N}, 0, \frac{\gamma^2 h}{N + 1} \mathcal{I}_2)$
                \For{$2 \leq k \leq N$}
                    \State $L_{i,j} \gets L_{i,j} * \phi(x_{i,j,k} - x_{i,j,k - 1} - \frac{\gamma^2 h}{N + 1} g_{i,j,k-1}, 0,  \frac{\gamma^2 h}{N + 1} \mathcal{I}_2)$
            \EndFor
            \State $L_{i,j} \gets L_{i,j} / q(B_{j,1},...,B_{j,N})$
            \State $L_i \gets L_i + L_{i,j}$
        \EndFor
        \State $L_i \gets L_i / M$
        \State $\ell \gets \ell + \log(L_i)$
      \EndFor
      \State \Return $\ell$
    \EndProcedure
  \end{algorithmic}
\end{algorithm}

\section{Numerical experiment} \label{sec:numexp}

\begin{figure}[H]
    \centering
    \includegraphics[width=0.49\linewidth]{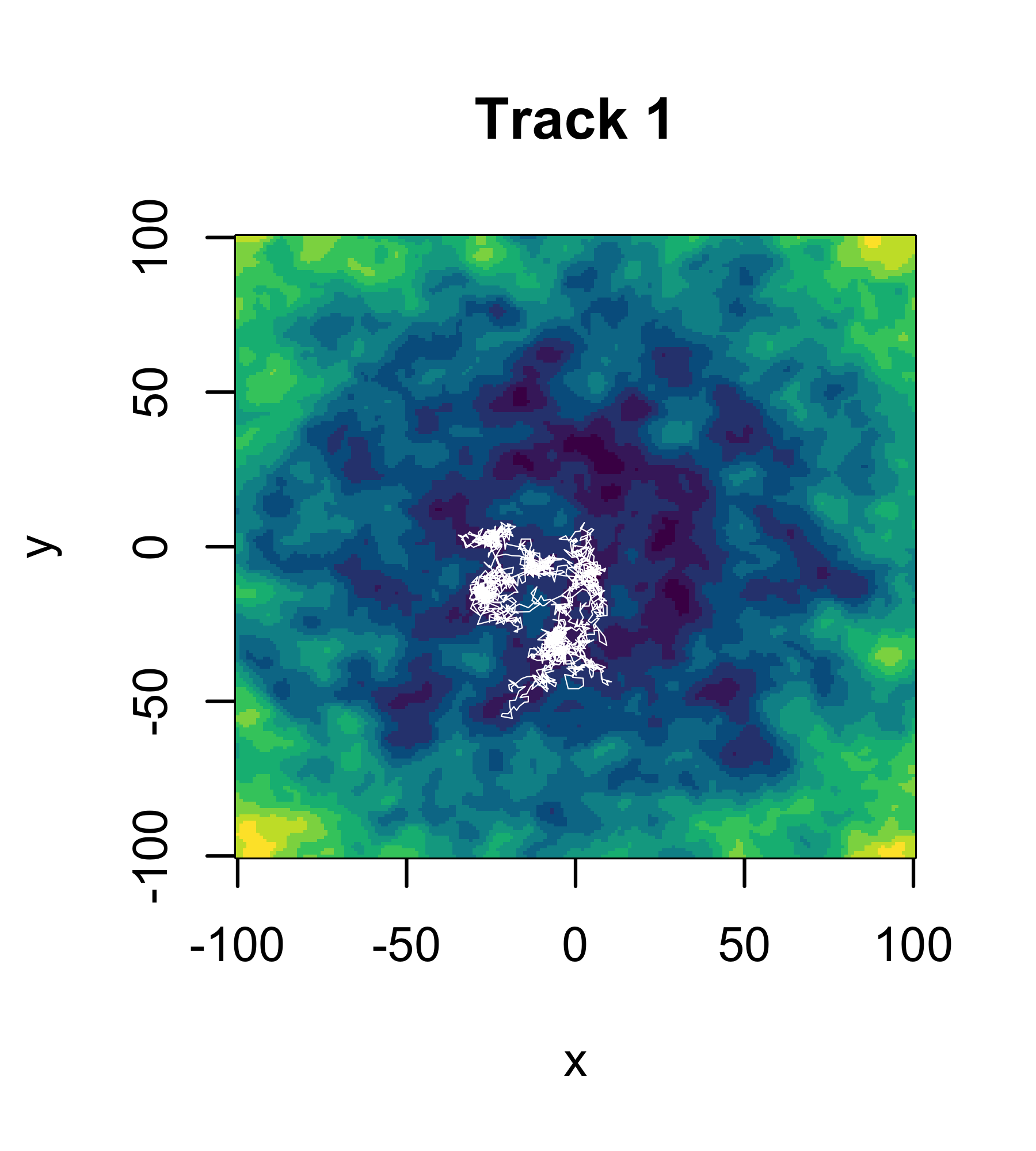}
    \includegraphics[width=0.49\linewidth]{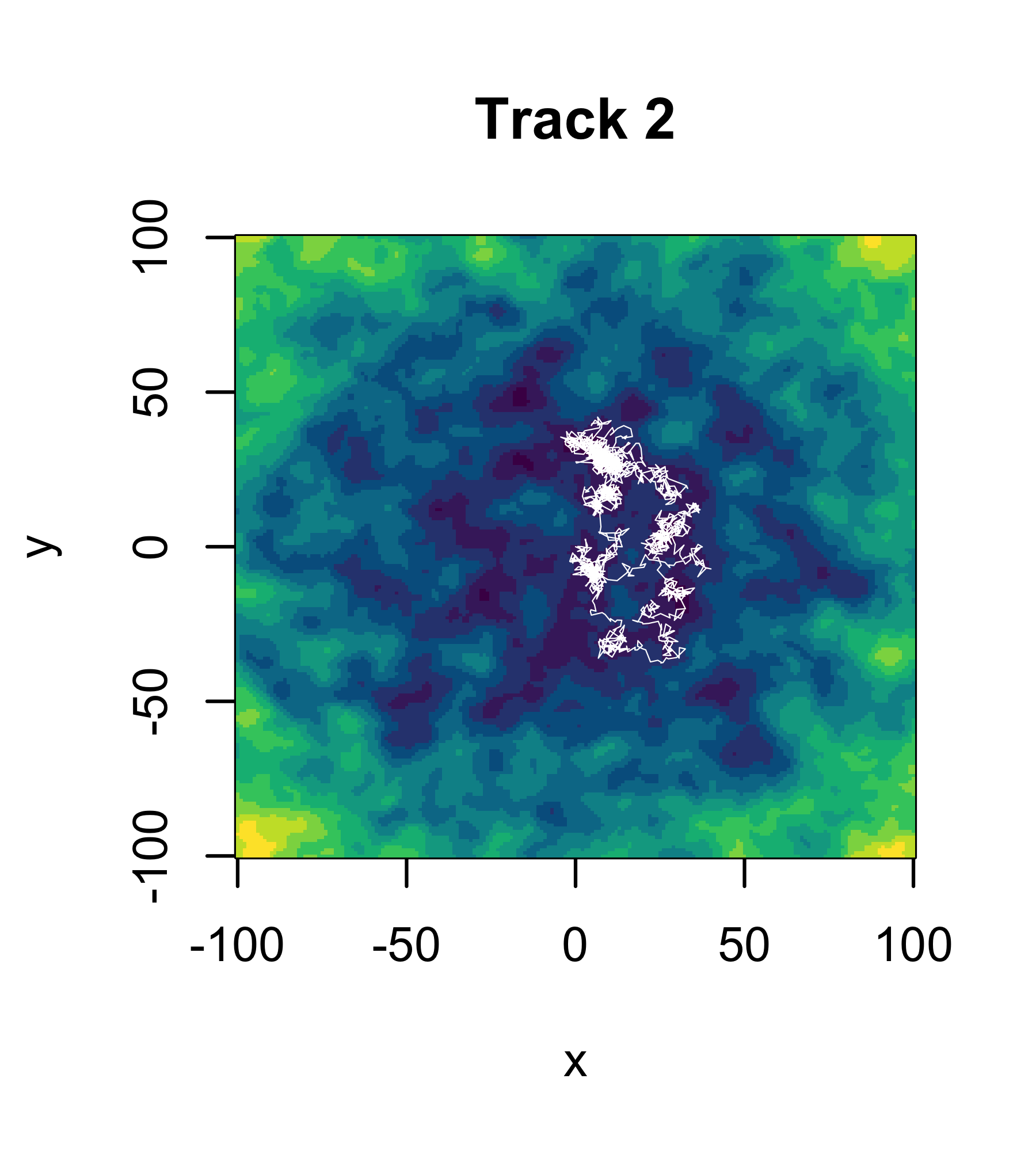}
    \caption{Two examples of simulated tracks from (\ref{eq:langevin-diffusion}), superimposed on the (log) RSF. $n = 2000$ measurements are displayed with a step size $\Delta_t = 0.5$ between measurements. Covariates were generated using Perlin noise. Blue corresponds to large values of the (log) RSF, yellow corresponds to smaller values.}
    \label{fig:tracks}
\end{figure}

We demonstrate the efficacy of the BBIS likelihood approximation by conducting four numerical experiments. All experiments were implemented in R. In each experiment, we simulate a large number of tracks from (\ref{eq:langevin-diffusion}) using an Euler-Maruyama approximation using a step-size of $0.01$ on the domain $[-100,100]^2$. We constructed the RSF $\pi$ using three covariate fields, giving us three covariate parameters $\Vec{\beta} = (\beta_1, \beta_2, \beta_3)$, and $\gamma^2$ to infer. These were fixed to $\Vec{\beta} = (4, 2, -0.1)$ and $\gamma^2 = 5$. The first two covariate fields were generated using the R-function \verb|noise_perlin| from the \verb|ambient|-package with  specified \verb|frequency = 0.05| and using the default values for the other parameters. The third covariate is the squared Euclidean distance to the center of the map.  We then created a simulated data set by thinning these tracks down to a rougher step size $\Delta_t$ and then attempted to recover parameters by maximising the BBIS-approximated log-likelihood function of the thinned tracks using \verb|optim|. Two examples of artificial animal tracks simulated using this method can be seen in Figure \ref{fig:tracks}, alongside the simulated log-RSF.

In the first two experiments we study the effect of step size $\Delta_t$ on the bias in the parameter estimates. This approach is based on the simulation studies conducted by Michelot et. al. \cite{Michelot2019}. The difference between the first two experiments is the method of thinning. In the first simulation study, for each $\Delta_t$ we consider, we keep number of measurements $n$ constant. The consequence of this is that the measurement span an increasingly larger amount of the simulated tracks as $\Delta_t$ increases. The the second simulation study, we instead keep constant the time $T$ spanned by the measurements. Thus the number of measurements $n$ decreases as $\Delta_t$ increases. In both simulation studies, the number of BBIS importance sampling nodes $N$ is selected so that the step size $h$ between nodes is $0.01$. The number of bridges is $M = 50$ for all simulations.

In the third and fourth experiment we aim to study the accuracy of the BBIS approximation as we vary the number of bridges $M$ and the number of importance sampling nodes $N$. We therefore fix $\Delta_t = 1$, and fix the number of observations to $n = 5000$. In the third experiment we fix $M = 50$ and vary $N$. In the fourth experiment we fix $N = 50$, and vary $M$.

For quick reference, we give a list with a summary of each of the four numerical experiments. We denote the first two simulation studies and the latter two convergence analysises.

\begin{itemize}
    \item \textbf{Simulation study I:} The number of observations is fixed. The time spanned by the observations increase as $\Delta_t$ increase.
    \item \textbf{Simulation study II:} Observations spans fixed time interval. Number of observations decrease as $\Delta_t$ increase.
    \item \textbf{Convergence analysis I:} $\Delta_t = 1$, the number of bridges $M = 50$, and the number of observations $n = 5000$ is fixed. The number of importance sampling nodes $N$ is varied.
    \item \textbf{Convergence analysis II:} $\Delta_t = 1$, the number of importance sampling nodes $N = 50$, and the number of observations $n = 5000$ is fixed. The number of bridges $M$ is varied.
\end{itemize}

\subsection{Simulation study I}
\begin{figure}[H]
    \centering
    \includegraphics[width=\linewidth]{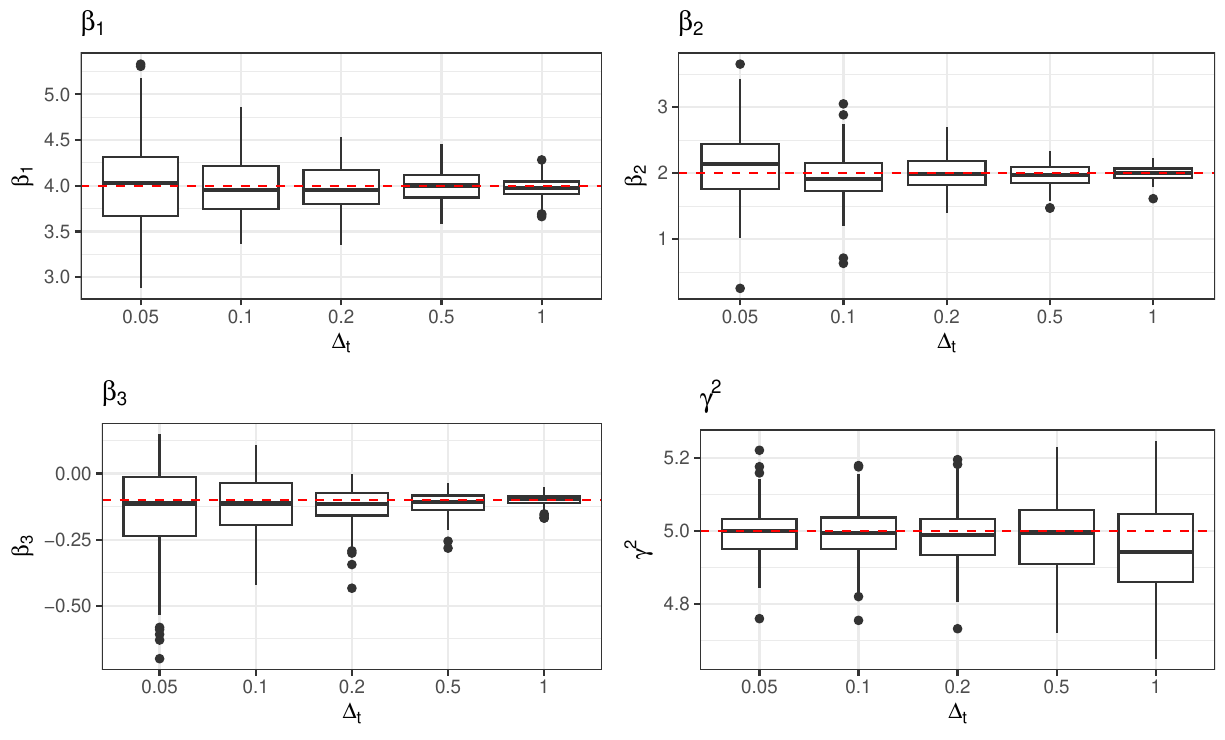}
    \caption[sim1]{Estimates of $\Vec{\beta}$ and $\gamma$ using $\Delta_t$ as distance between observations and 5000 observations. The red dotted line shows the true value of the parameters}
    \label{fig:sim1}
\end{figure}

In the first numerical experiment we simulated 100 tracks with step-size $0.01$ and thinned them to each of the step-sizes $\Delta_t \in  \{0.05, \ 0.1, \ 0.2, \ 0.5, \ 1\}$, with $n = 5000$ observations per track. For each of the tracks we estimated the parameters $\Vec{\beta}$ and $\gamma^2$ using the BBIS likelihood approximation. We used $M=50$ bridges and a number of nodes such that the importance sampling step size $h$ was $0.01$. The resulting estimates of $\Vec{\beta}$ and $\gamma$ are shown in Figure~\ref{fig:sim1}.

%\textcite{Michelot2019} found that there was decreasing variance in the estimates of $\Vec{\beta}$ as the value of $\Delta_t$ increases.
We observe in Figure~\ref{fig:sim1} that the variance decreases when $\Delta_t$ increases; this can be explained by the fact that increasing $\Delta_t$ while keeping the number of observation constant increases the time spanned by the measurements and therefore makes them explore more of the RSF. Figure~\ref{fig:sim1} also shows that there is no considerable bias present in the estimates for any of the values of $\Delta_t$ tested.

\subsection{Simulation study II}
\begin{figure}[H]
    \centering
    \includegraphics[width=\linewidth]{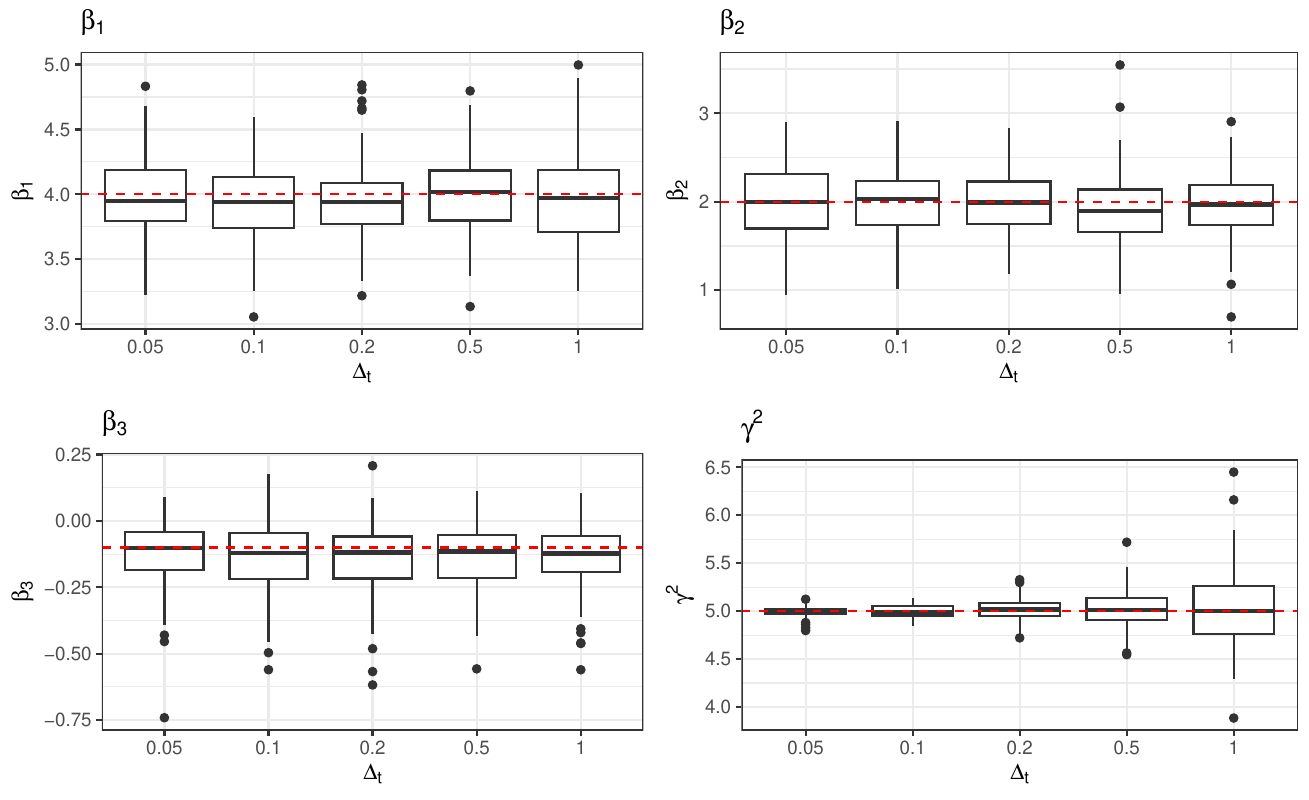}
    \caption[sim2]{Estimates of $\Vec{\beta}$ and $\gamma$ using $\Delta_t$ as distance between observations and 500 as time span of tracks. The red dotted line shows the true value of the parameters}
    \label{fig:sim2}
\end{figure}

In the second numerical experiment we again simulated $100$ tracks with step-size $0.01$ and thinned them to each of the step-sizes $\Delta_t \in  \{0.05, \ 0.1, \ 0.2, \ 0.5, \ 1\}$, but now we use only the measurements where $t \leq 500$. We used $M=50$ and a number of nodes such that the importance sampling step size $h$ was $0.01$. The resulting estimates are shown in Figure~\ref{fig:sim2}.

From Figure~\ref{fig:sim2} we see that when the time spanned by the tracks is held constant, the variance of the $\beta$ estimates is similar for all the tried values of $\Delta_t$. On the other hand there is a large increase in the variance of the $\gamma^2$ estimates. This might be explained by the fact that, when we increase $\Delta_t$ while keeping the time spanned by the tracks constant, the number of observations go down.

\subsection{Convergence analysis I}

\begin{figure}[H]
    \centering
    \includegraphics[width=\linewidth]{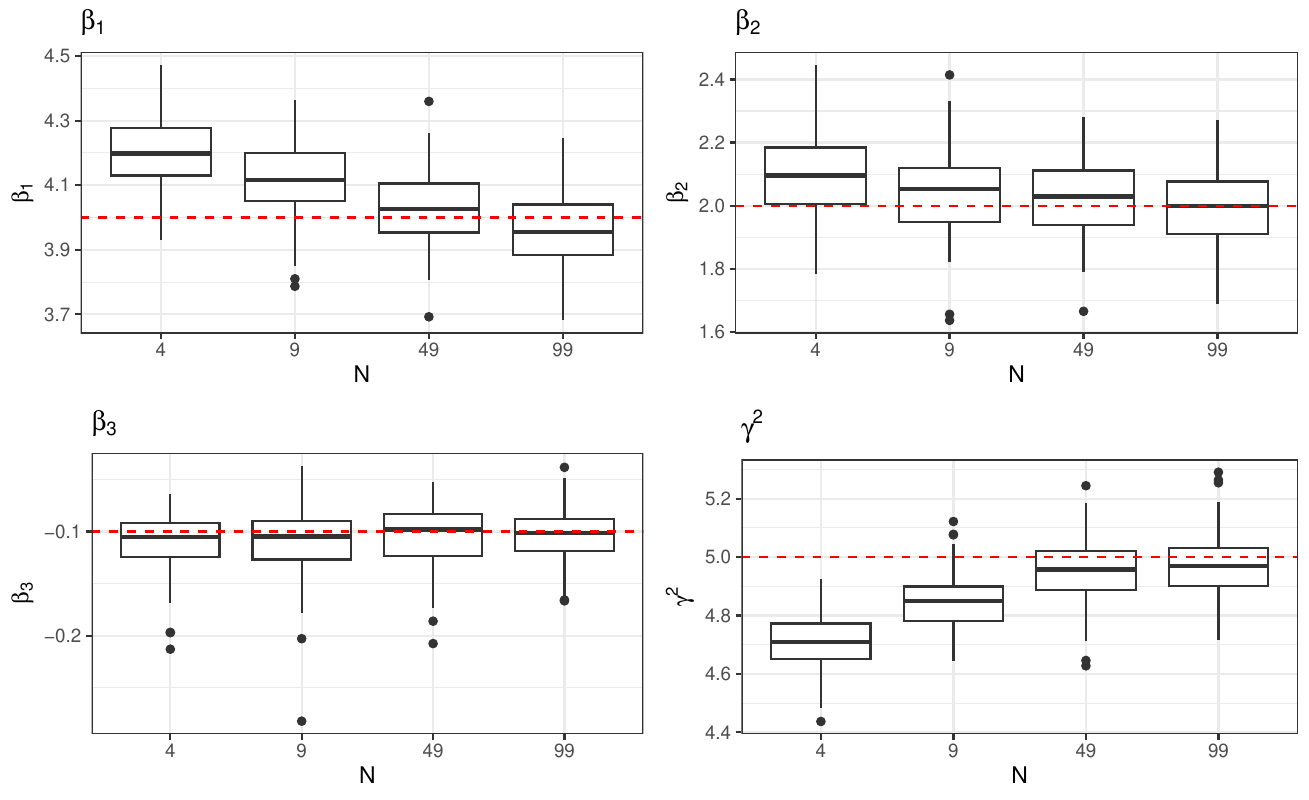}
    \caption[sim3]{Estimates of $\Vec{\beta}$ and $\gamma$ using $N$ bridge nodes, $M=50$ bridges, $\Delta_t=1$ and 5000 observations. The red dotted line shows the true value of the parameters}
    \label{fig:sim3}
\end{figure}

In the third numerical experiment we test how the number of importance sampling nodes $N$ affects the bias in parameter estimates. We did this by simulating $100$ tracks and then thinning them by a factor of $100$ to get $\Delta_t = 1$. For each of these tracks we estimated the parameters $\Vec{\beta}$ and $\tau^2$ using $M=50$ bridges and $N =\{4,9,49,99\}$ importance sampling nodes. Boxplots of the resulting estimates are shown in Figure~\ref{fig:sim3}.

Figure~\ref{fig:sim3} shows that there is a bias in the estimates of $\beta$ and $\gamma^2$ for small values of $N$, but this bias disappears as the resolution of the bridges becomes similar to that of the simulations.

\subsection{Convergence analysis II}
In the fourth numerical experiment we test how the number of bridges $M$ affects the bias in parameter estimates. We did this by simulating $100$ tracks and then thinning them by a factor of $100$ to get $\Delta_t = 1$. For each of these tracks we estimated the parameters $\Vec{\beta}$ and $\tau^2$ using $N=50$ importance sampling nodes and $M \in \{5 , \ 10, \ 50, \ 100, \ 200\}$ bridges. Boxplots of the resulting estimates are shown in Figure~\ref{fig:sim3}.

\begin{figure}[H]
    \centering
    \includegraphics[width=\linewidth]{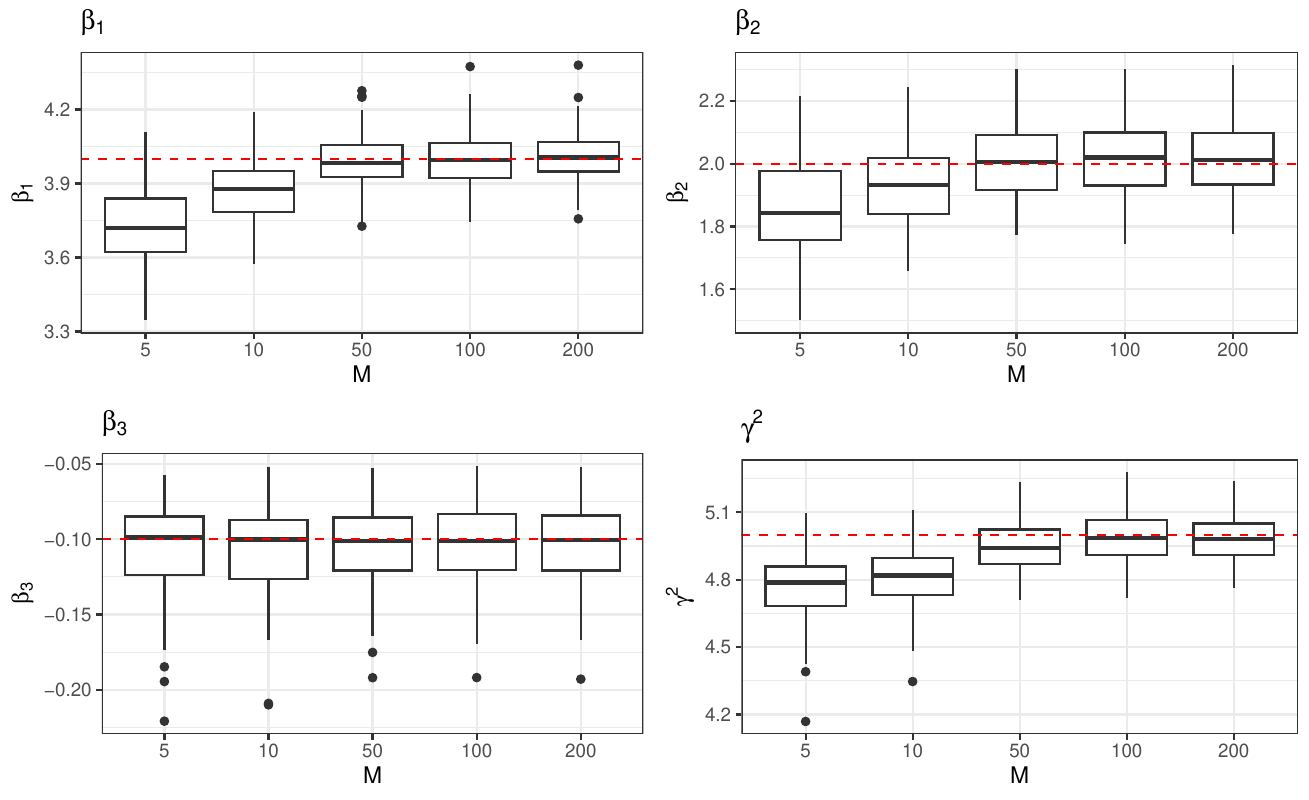}
    \caption[sim4]{Estimates of $\Vec{\beta}$ and $\gamma$ using $N=50$ bridge nodes, $M$ bridges, $\Delta_t=1$ and 5000 observations. The red dotted line shows the true value of the parameters}
    \label{fig:sim4}
\end{figure}

In Figure~\ref{fig:sim4} we see that if the number of bridges is too low, there is a bias in the parameter estimates and the spread of the estimates is slightly larger. At $M=50$ bridges none of the estimates show a bias, and increasing the number of bridges does not yield any noticable improvements.

\subsection{Computation time and memory}

\begin{figure}[H]
    \centering
    \includegraphics[width=0.45\linewidth]{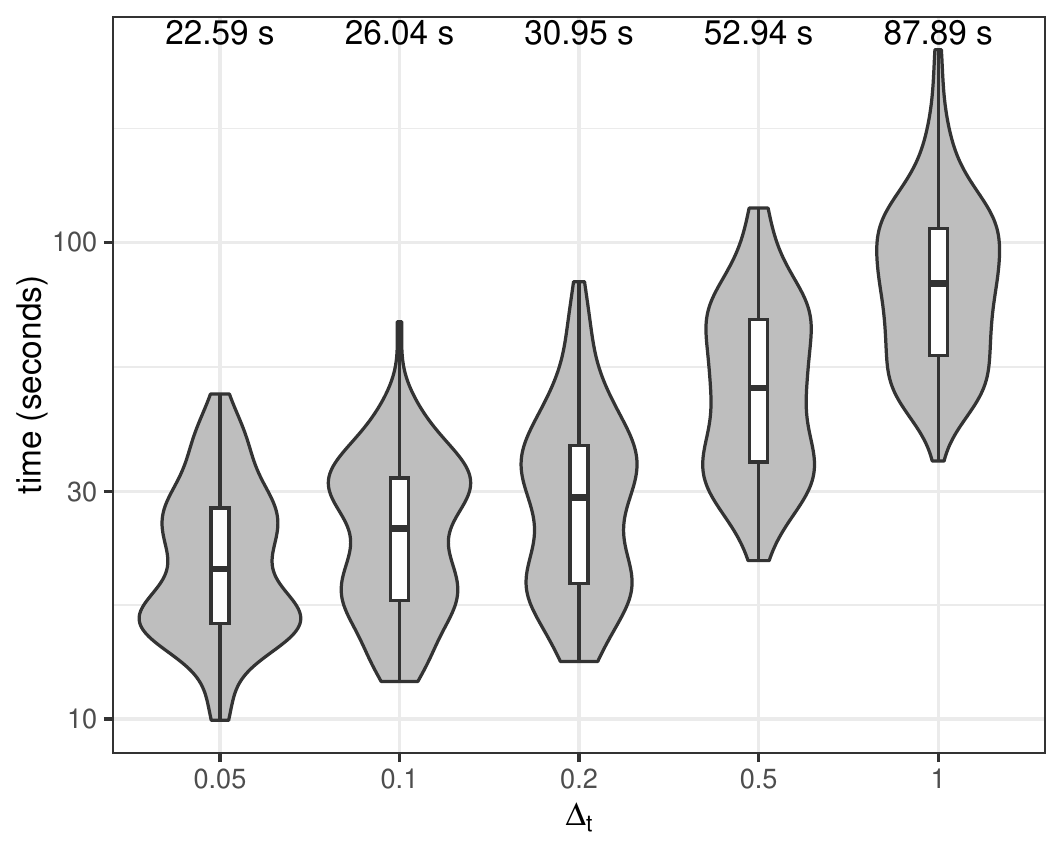}
    \includegraphics[width=0.45\linewidth]{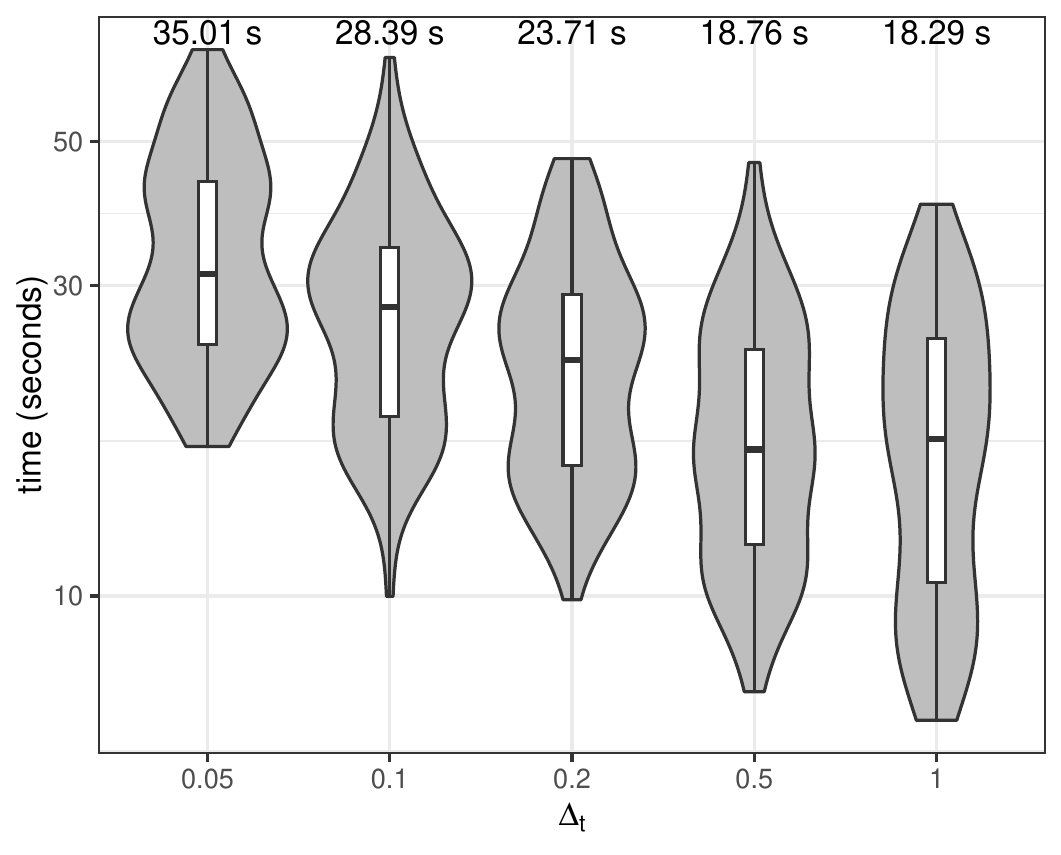}
    \caption[comp-time-1]{Violin plots of the computation time used in simulation study I (left) and simulation study II (right). The median compute time for the experiment is reported at the top of each violin.}
    \label{fig:comp-time-1}
\end{figure}

\begin{figure}[H]
    \centering
    \includegraphics[width=0.45\linewidth]{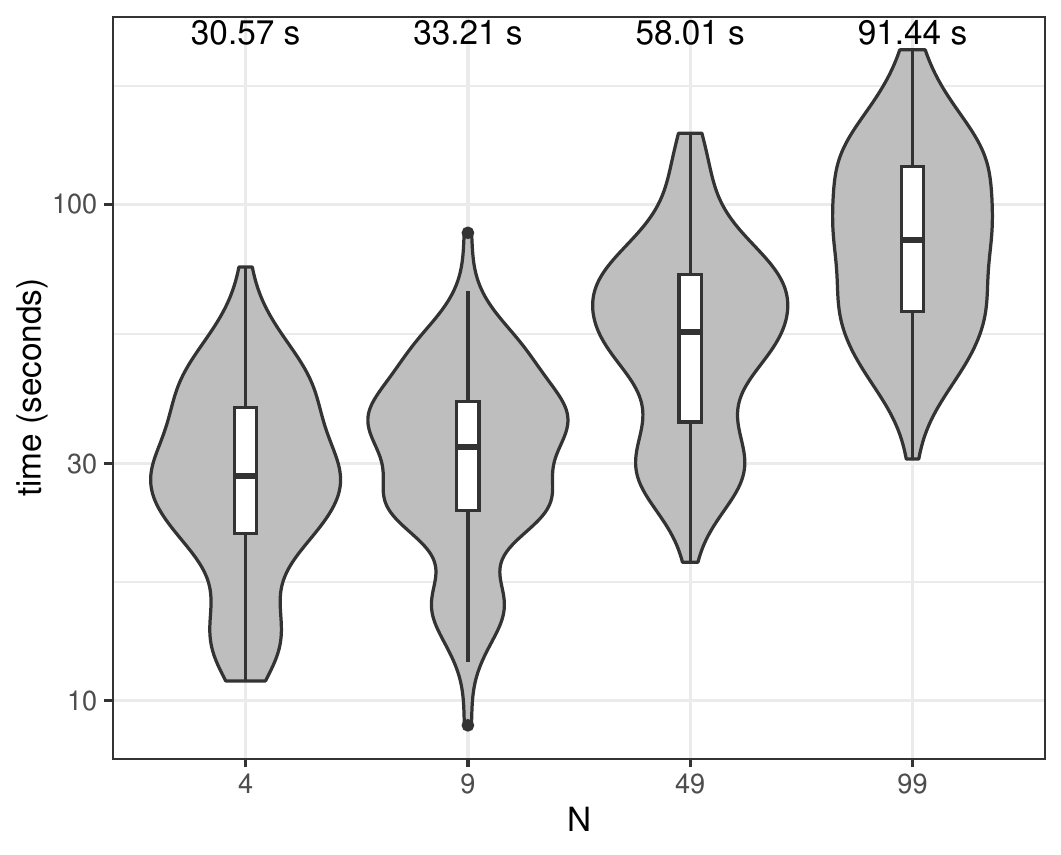}
    \includegraphics[width=0.45\linewidth]{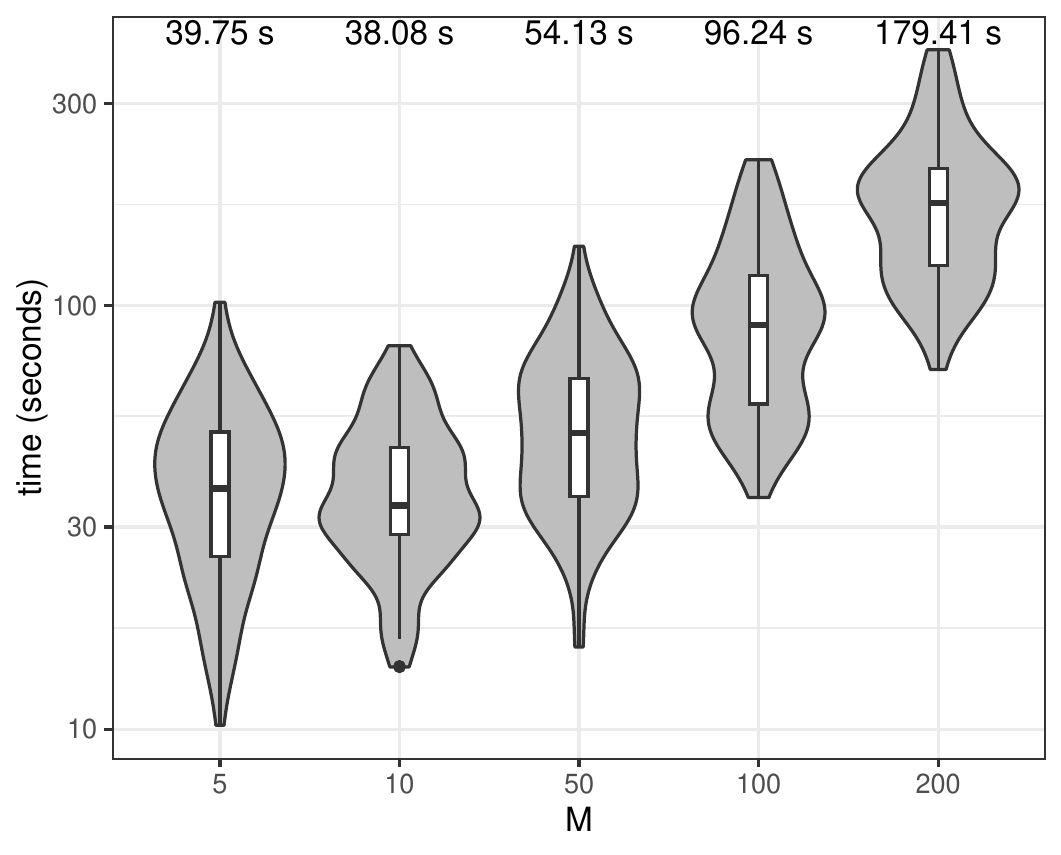}
    \caption[comp-time-2]{Violin plots of the computation time used in convergence analysis I (left) and convergence analysis II (right). The median compute time for the experiment is reported at the top of each violin.}
    \label{fig:comp-time-2}
\end{figure}

One weakness of the BBIS method vis a vis Euler is the increased computation time. Adding additional nodes to our discretisation naturally increases the computational cost of inference. However, the computation times are still very manageable. Violin plots of the computations times in our numerical experiments can be seen Figure \ref{fig:comp-time-1} and Figure \ref{fig:comp-time-2}. With $M = 200$, $N = 50$, and $5000$ observations, we still have a computation times $<2$ minutes, which is the largest among all of our numerical experiments. Another weakness of our method is the memory requirement; if a very large number of nodes and bridges are required for accurate approximation, then storing all the pre-computed bridges can be infeasible.

\section{Case study}

\begin{figure}[H]
    \centering
    \includegraphics[width=\linewidth]{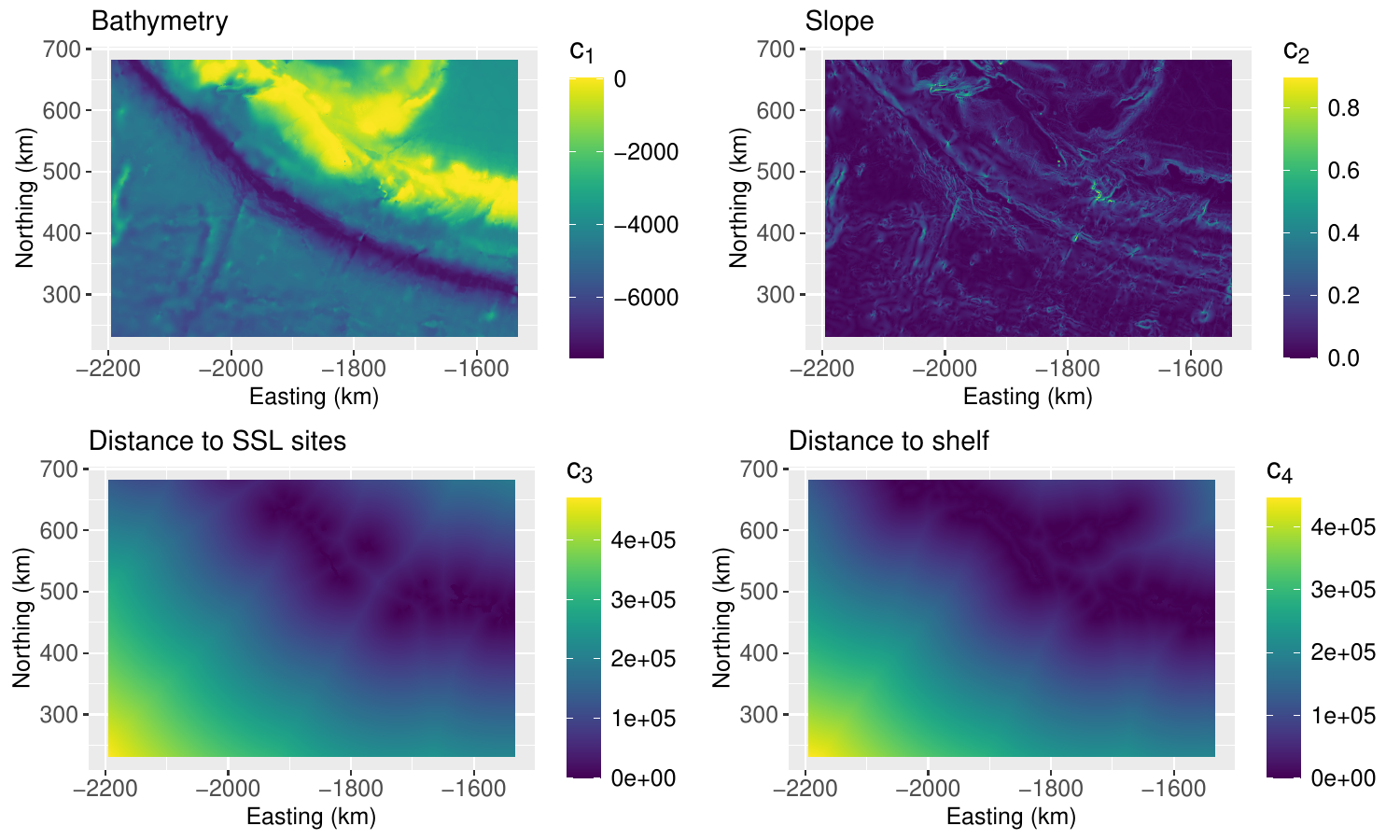}
    \caption[sim4]{Covariates used for Steller sea lion analysis}
    \label{fig:covariates}
\end{figure}

We replicated fitting the Langevin process from Michelot et al. \cite{Michelot2019} based on the Steller sea lion data set described in Wilson et al. \cite{wilson2018estimating}. The data set is comprised of the trajectories of three individuals collected on irregular intervals using the Argos satellite network, in addition to four spatial covariates: bathymetry ($c_1$), slope ($c_2$), distance to seal haulout sites ($c_3$), and distance to continental shelf($c_4$) (Figure \ref{fig:covariates}). Due to the high correlation between $c_3$ and $c_4$, as in Michelot et al. \cite{Michelot2019}, we excluded $c_4$ from the analysis. Argos Locations can have high spatial error, which we corrected, as in Michelot et al. \cite{Michelot2019}, using the R package \verb|crawl| (Johnson and London \cite{johnson2018crawl}). The \verb|crawl| package uses a Kalman filter to fit a continuous-time Brownian motion model to the Argos locations in order to estimate the true locations of the trajectories at the observation times.

In order to generalise the model to irregularly sampled data, we defined a maximum time interval $\Delta_{\max}$ between nodes and set the number of nodes to be  $N = \left\lceil \frac{\Delta_i}{\Delta_{\max}} \right\rceil - 1$, where $\Delta_i$ was the interval between observations $i$ and $i+1$. To test the effect of temporal resolution of integration nodes, we considered 30 values for $\Delta_{max}$ that were regularly spaced between 0.01 and 25 on the log scale. For each of the values of $\Delta_{max}$ the BBIS method was tested 10 times for each of the numbers of bridges $M \in {25, 50, 100}$, in order to explore the convergence and spread of the estimates.

\subsection{Results}
\begin{figure}[H]
    \centering
    \includegraphics[width=\linewidth]{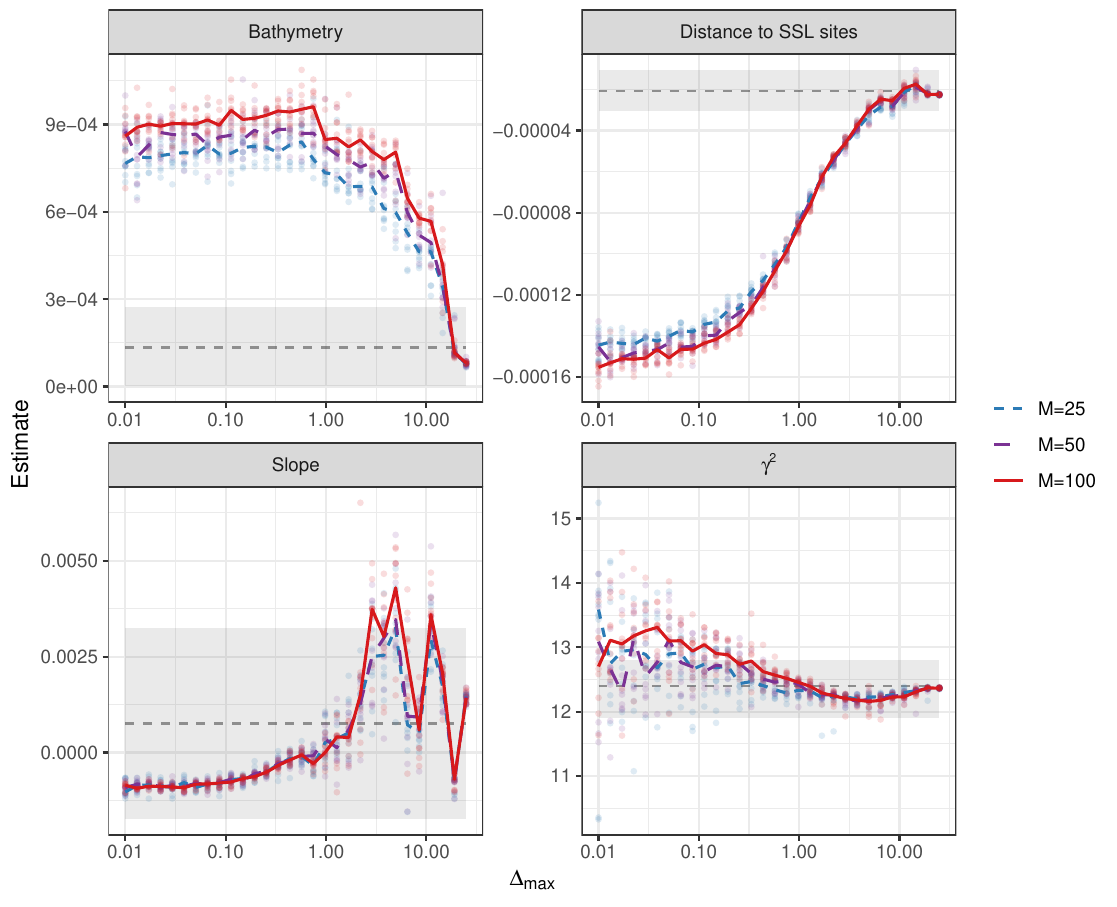}
    \caption[sim4]{Estimates of $\Vec{\beta}$ and $\gamma$ against a grid of values for $\Delta_{max}$ on log-scale. The grey dotted line and the grey ribbon show the estimates and confidence interval found by Michelot et al. \cite{Michelot2019}}
    \label{fig:SSL_estimates}
\end{figure}

The results of the coefficient estimates with respect to the selected values of $\Delta_{max}$ and $M$ are shown in Figure~\ref{fig:SSL_estimates}. We observed that coefficient estimates were very similar to Michelot et al. \cite{Michelot2019} at large values of $\Delta_{max}$ (i.e., when BBIS was applied only  to large gaps). However, as $\Delta_{max}$ decreased, the coefficient estimates diverged outside of the confidence intervals estimated in Michelot et al. \cite{Michelot2019}. For bathymetry ($\beta_1$), estimates stabilised at $\Delta_{max}<1.0$ hours, and estimates for distance to SSL haulout sites ($\beta_2$) stabilised at $\Delta_{max}<0.1$ hours. Interestingly, when $\Delta_{max}>1.0$ hours, there is a lot of noise in the parameter estimates for slope of bathymetry ($\beta_2$), however, the estimates stabilised when $\Delta_{max}<0.1$ hours. Notably, there was a sign flip compared to the estimate of Michelot et al. \cite{Michelot2019} that occurred round $\Delta_{max} = 1$. Coefficient estimate for distance to haul sites ($\beta_3$) are similar to bathymetry, with coefficient estimates becoming more extreme as $\Delta_{max}$ decreased, and stabilising when $\Delta_{max}<0.1$.

\begin{figure}[H]
    \centering
    \includegraphics[width=\linewidth]{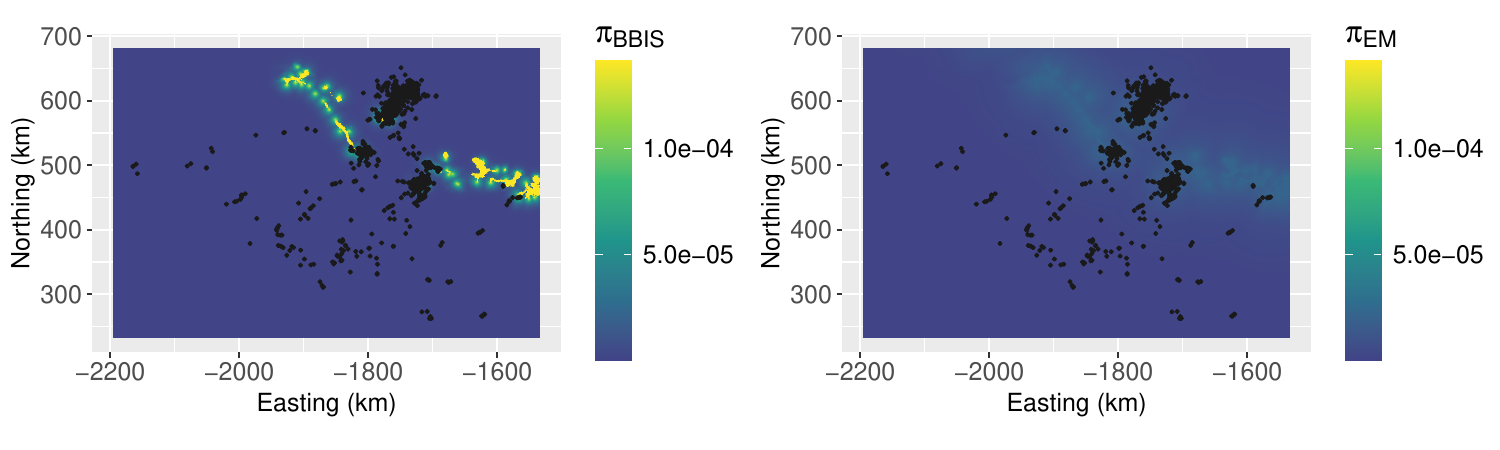}
    \includegraphics[width=0.5\linewidth]{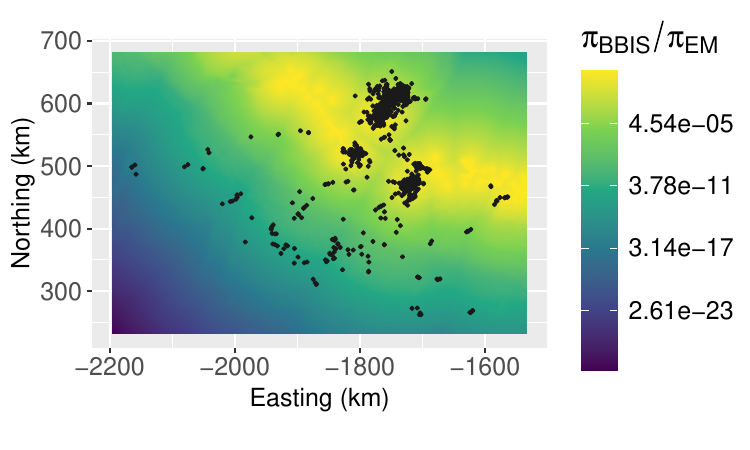}
    \caption[sim4]{Estimates of the utilization distribution using estimates of $\beta$ from the BBIS likelihood (assuming $\Delta_{max}=0.01$ and $M=100$) (upper left), estimates from Michelot et al. \cite{Michelot2019} (upper right) and the ratio of the first UD over the second UD(lower). The black dots represent filtered Steller sea lion locations.}
    \label{fig:SSL_UDs}
\end{figure}

For slope and distance to haulout sites, the coefficients seem to have converged with respect to the number of bridges. For the diffusion parameter $\gamma^2$ and bathymetry, there are some difference in the value of estimates depending on number of bridges at the lower end of values of $\Delta_{max}$. This suggests that more bridges may be necessary in order to obtain the true MLE parameters of the Langevin model. The fact that the magnitude of the coefficients for bathymetry and distance to SSL sites is larger using the BBIS likelihood is consistent with the finding in Michelot et al. \cite{Michelot2019}, that the Euler-Maruyama method under-estimates the coefficients. The effect of under-estimation of coefficients is also apparent in the estimated UD. Specifically, the UD estimated using the BBIS likelihood is more concentrated than the estimate from Michelot et al \cite{Michelot2019} (Figure \ref{fig:SSL_UDs}).

\section{Discussion}

In this work we have built on the analysis of the Langevin diffusion as a model for animal movement by suggesting a new method for performing statistical inference. As described in Section \ref{sec:bbis} the BBIS method uses Brownian bridges in an importance sampling integrator to provide an approximation of the log-likelihood function that is more accurate and precise than the Euler-Maruyama approximation. Furthermore, the BBIS method can be made arbitrarily accurate at the cost of additional compute and memory by increasing the number of importance sampling nodes $N$ and the number of Brownian bridges $M$. As observed in Section \ref{sec:numexp} the biases in parameter estimates observed by Michelot, et. al. \cite{Michelot2019} are almost completely eliminated when $M$ and $N$ are both sufficiently large. Importantly, a large reduction in the bias can be achieved using moderate values for $M$ and $N$. This suggests that the bias observed by Michelot et al. is not an inherent feature of the Langevin diffusion model, but a consequence of the approximation error in the Euler-Maruyama scheme for large time-steps. 

In Simulation study I we observed that the variance of the estimates for the $\beta$ coefficients was greatly reduced when we increased the time spanned by the tracks, while in Simulation study II we observed that the variance was largely independent of the frequency of locations (i.e., increased number of observations over the same duration). This would suggest that to obtain accurate estimates for the RSF, the time spanned by the observations is more important than the frequency of locations. Therefore, if you can only make a fixed number of measurements, the best strategy is to spread them over as long a time period as possible. With respect to the estimation of $\gamma^2$, Simulation study I showed that the variance in the estimates increased with the time spanned, while in Simulation study II we observed that the variance decreased at higher location frequency. This suggests that the best strategy for estimating $\gamma^2$ is to make the sampling frequency as large as possible. Given that animal tracking tags have limited batteries, the choice of sampling regime is typically a trade-off between sampling frequency and sampling duration \cite{he2023guide}. Simulations I and II, therefore revealed that this trade-off may subsequently affect the precision in the estimation of the $\beta$s (which favours sampling duration over frequency) and $\gamma^2$ (which favours frequency over duration) coefficients. Luckily, it seems that it is possible to get accurate estimates for the $\beta$'s, even if the uncertainty in the estimate of $\gamma^2$ is large, and vice versa.

The main weakness of our approach is the increased compute time. Running statistical inference with the BBIS likelihood approximation with sufficient $M$ and $N$ takes much longer than a comparative inference using an Euler-Maruyama approximation. As described in Section \ref{sec:bbis}, we used the same Brownian bridges in each likelihood evaluation in order to improve the stability of gradient-based optimisation and to reduce computation of re-simulating the bridges. Storing all the Brownian bridges can be very memory-intensive, which may make our implementation intractable for very large datasets. Additionally, the BBIS method is not in itself capable of handling data with measurement error.

There are many avenues for potentially improving the BBIS method. Durham \& Gallant \cite{Durham2002} discussed multiple alternative, and possibly more efficient, ways to construct proposal bridges between measurements. Their modified Brownian bridge more aggressively shrinks the proposal variance as paths approach the observed endpoint; a minor modification to the Brownian bridge that they claimed significantly increased efficiency. Subsequent work claims to deliver even more efficient schemes for bridge construction \cite{Lindstrm2011, Whitaker2016}. Durham \& Gallant \cite{Durham2002} also discussed using higher order approximations of the transition density in (\ref{eq:monte-carlo-approximation}), most notably the Ozaki scheme. Also, it should be possible to further augment these methods with techniques from the multi-level Monte-Carlo literature (see \cite{Heinrich2001} for a review). There also exist other approaches in the literature for inferring parameters in discretely sampled Itô diffusion models, most notably the method by Aït-Sahalia \cite{AitSahalia2008, AitSahalia2002} based on Hermite polynomial approximations, and moment estimators based on the method of estimating functions (see \cite{Bibby2010} for a review). Finally, it might be possible to deal with measurement error by combining the BBIS method with a particle filter.

We demonstrated the BBIS method can reduce bias in estimates of the habitat selection coefficients. In addition, we observed that sampling duration was more important than sampling frequency in improving estimates of habitat coefficient. This suggests that the method may be well suited for modelling coarser tracking data, which is particularly common with older generations of animal tags \cite{he2023guide}. Furthermore, the continuous-time nature of the Langevin diffusion model means that it is possible to jointly model data with different sampling regimes. Integrating data from both coarse and fine sampling regimes should improve the estimates of both habitat and movement coefficients. In addition to integrating existing data, our method may provide future research with an incentive to use a coarser and longer sampling regime. By increasing the accuracy of habitat selection estimates as well as the breadth of data that can be modelled, our method should enable improved ecological inference into habitat selection. 

\section{Declarations}
\noindent
\textbf{Funding.} RRT is supported by the Research Council of Norway (GreenPlan, Project No. 326979).

\noindent
\textbf{Conflicts of interest.} \, The authors declare no conflicts of interest.

\newpage
\printbibliography

\end{document}